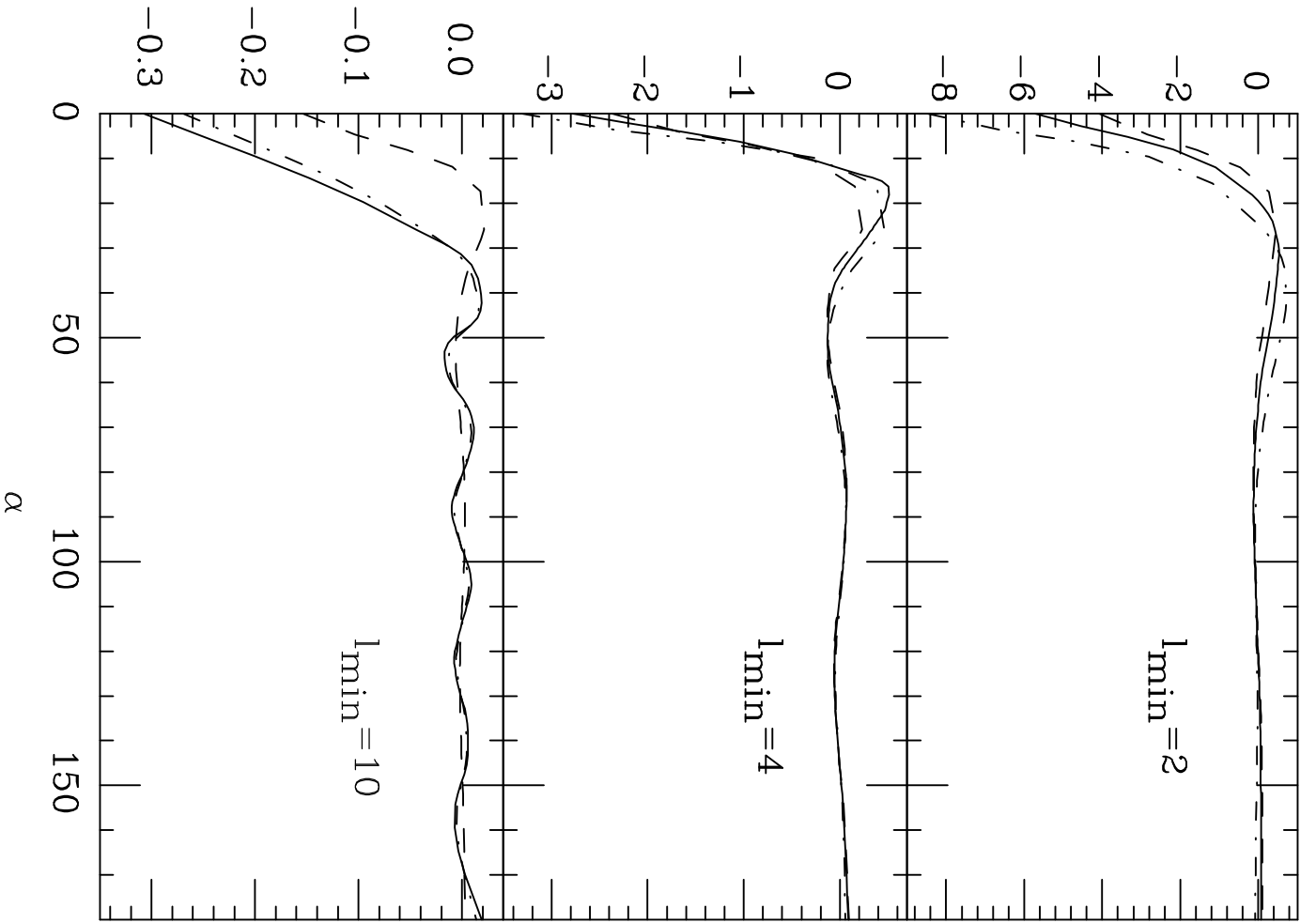

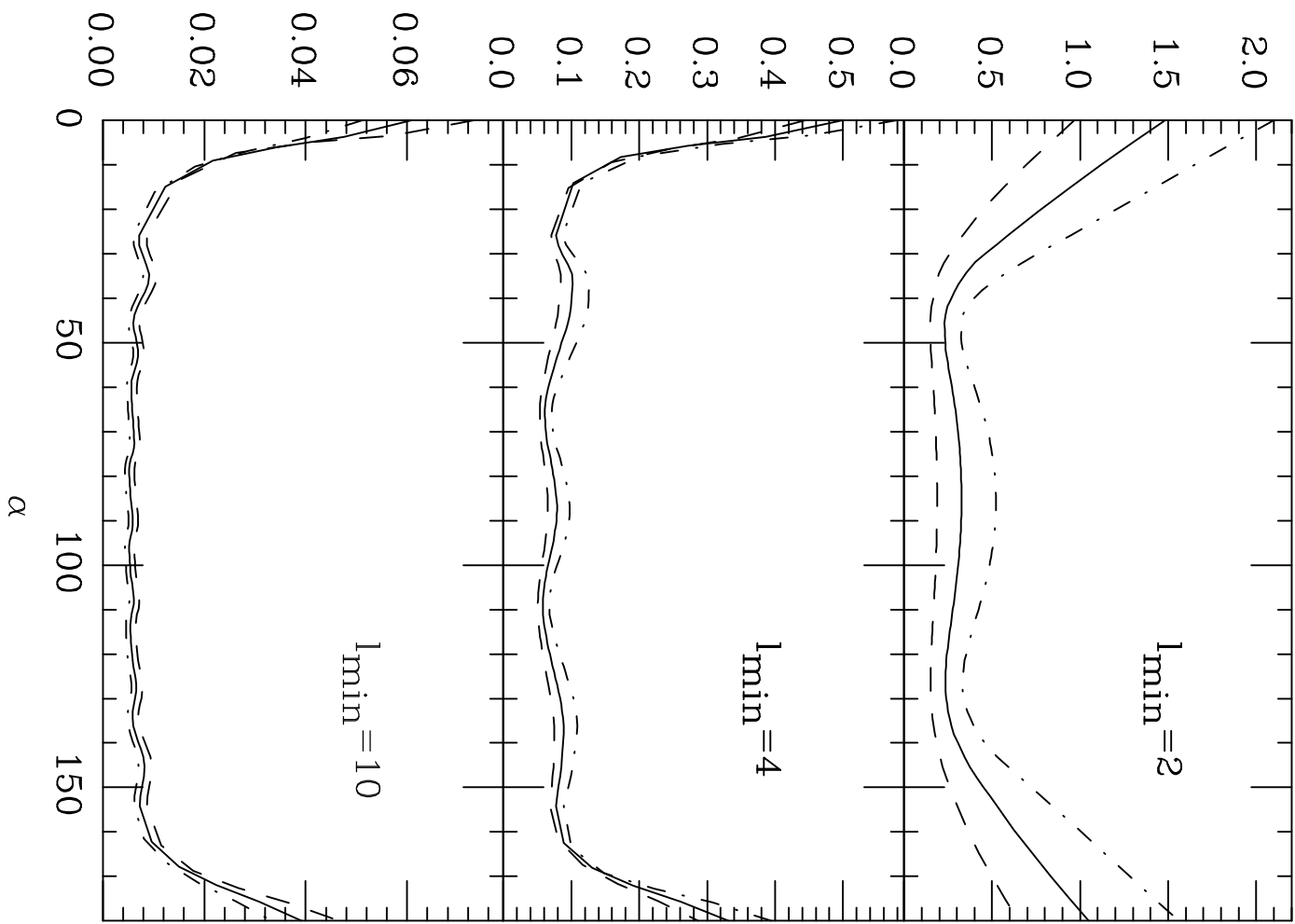

# Contribution to the Three–Point Function of the Cosmic Microwave Background from the Rees–Sciama Effect

**Silvia Mollerach**[1], **Alejandro Gangui**[1],
**Francesco Lucchin**[2] and **Sabino Matarrese**[3]

[1]SISSA – International School for Advanced Studies,
via Beirut 2–4, I–34013 Trieste, Italy

[2]Dipartimento di Astronomia, Università di Padova,
vicolo dell'Osservatorio 5, I–35122 Padova, Italy

[3]Dipartimento di Fisica G. Galilei, Università di Padova,
via Marzolo 8, I–35131 Padova, Italy

**Abstract**

We compute the contribution to the three–point temperature correlation function of the Cosmic Microwave Background coming from the non–linear evolution of Gaussian initial perturbations, as described by the Rees–Sciama (or integrated Sachs–Wolfe) effect. By expressing the collapsed three–point function in terms of multipole amplitudes, we are able to calculate its expectation value for any power spectrum and for any experimental setting on large angular scales. We also give an analytical expression for the *rms* collapsed three–point function arising from the cosmic variance of a Gaussian fluctuation field. In the case of *COBE* DMR, we find that the predicted signal is about three orders of magnitude below that expected from the cosmic variance.

After the detection by *COBE* DMR (e.g. Smoot et al. 1992, Bennett et al. 1994) of anisotropies in the Cosmic Microwave Background (CMB), several other measurements of anisotropies at different angular scales have been announced. This new set of observations has provided an extra tool to improve our understanding of the structure formation process. The usual analysis in terms of the two–point correlation function of the *COBE* data fixes the *rms* amplitude of the primordial perturbations. A comparison with measurements at smaller angular scales encodes information about the matter content of the universe and other cosmological parameters to which these are sensitive. Furthermore, an analysis of the three–point correlation function of the data, as it has recently been performed for the *COBE* DMR first and two–year data (Hinshaw et al. 1994, 1995), can provide useful clues about the statistical properties of cosmological perturbations. Further analyses probing the statistical nature of temperature fluctuations have been performed on the *COBE* maps by Smoot et al. (1994), Kogut et al. (1995) and Torres et al. (1995). These studies are of particular relevance as they could help to distinguish among the two presently preferred theories for the origin of primordial fluctuations: inflation and topological defects.

Inflationary models predict a quasi–Gaussian distribution of density perturbations. This is due to the fact that they are originated by the quantum fluctuations of a very weakly coupled scalar field, the *inflaton*. The effect of the small non–linearities in the inflaton dynamics and the mean value of the resulting imprints on the three–point function of the gravitational potential (see Falk, Rangarajan & Srednicki 1993, Gangui et al. 1994, Gangui 1994) are several orders of magnitude smaller than the typical values expected for a particular realization of an ensemble of Gaussian universes, the so–called cosmic variance (Scaramella & Vittorio 1991, Srednicki 1993). As the anisotropies observed at large angular scales are essentially determined by the fluctuations in the gravitational potential, the CMB three–point correlation function at the scales proved by *COBE* is also expected to be Gaussian. An analysis of the *rms* skewness of inflationary models giving rise to isocurvature baryon perturbations, by Yamamoto and Sasaki (1994), yields values smaller than those obtained for curvature perturbation models. On the other hand, topological defects are the typical example of non–Gaussian distributed perturbations. However, for many observations, the relevant object is not the individual effect of each particular defect, but the superposition of many of them, which results in a nearly Gaussian pattern (e.g. Gott et al. 1990,

Coulson et al. 1994, Gangui & Perivolaropoulos 1995). Moreover, it has been shown (Scherrer & Schaefer 1994) that, for a wide class of models leading to non–Gaussian density perturbations, the corresponding temperature fluctuation field, as induced by the Sachs–Wolfe effect, is nearly Gaussian thanks to the Central Limit Theorem. Thus, a detailed analysis of the predictions for various models, taking into account all the relevant effects, is necessary to find out the best tools to distinguish among them.

The analysis of the three–point correlations in the *COBE* DMR two–year anisotropy maps shows evidence for a non–vanishing signal in the data but at a level consistent with a superposition of instrumental noise and Gaussian CMB fluctuations (Hinshaw et al. 1995). As the noise level will diminish rapidly with additional data, in the four–year map the sensitivity is expected to be limited by the cosmic variance.

Even for the case of primordial Gaussian curvature fluctuations, the non–linear gravitational evolution gives rise to a non–vanishing three–point correlation function of the CMB. It has been argued by Luo & Schramm (1993) that the amplitude of this effect is several orders of magnitude larger than that predicted by Falk, Rangarajan & Srednicki (1993) for a cubic self–interacting inflaton model, and with a similar angular dependence. A more recent estimate of this effect on the skewness of the CMB by Munshi, Souradeep & Starobinsky (1995) finds that the amplitude of the gravitational and inflationary non–linearity contributions is comparable for typical inflation models. We provide here a detailed and more general analysis of the possible observational consequences of the non–linear gravitational growth of initially Gaussian perturbations on the CMB three–point function.

At large angular scales the anisotropies in the CMB are given by (e.g. Martínez–González, Sanz & Silk 1990)

$$\frac{\Delta T}{T}(\hat{\gamma}) = \frac{1}{3}\Phi(\hat{\gamma}\eta_0, \eta_r) + 2\int_{\eta_r}^{\eta_0} d\eta \frac{\partial}{\partial \eta}\Phi(\mathbf{x}, \eta)\bigg|_{\mathbf{x}=\hat{\gamma}(\eta_0-\eta)}, \qquad (1)$$

where the first term represents the well–known Sachs–Wolfe effect (Sachs & Wolfe 1967), while the second one corresponds to the Rees–Sciama, or integrated Sachs–Wolfe effect (Rees & Sciama 1968). In the previous formula $\hat{\gamma}$ denotes a direction in the sky and $\eta$ is the conformal time, with $\eta_0$ and $\eta_r$ the present and recombination times, respectively. The contribution of the Rees–Sciama effect to the total anisotropy is small in a flat matter–dominated universe because the gravitational potential keeps constant in time within the linear regime. It gives however a non–vanishing contribution when the non–linear evolution of perturbations is taken into account. This non–linear contribution will generate a non–vanishing three–point function, even for primordial Gaussian perturbations in the energy density. Second–order perturbation theory gives, for the flat matter–dominated case, $\Phi(\mathbf{x}, \eta) = \Phi_1(\mathbf{x}) + \Phi_2(\mathbf{x}, \eta)$, where the second–order gravitational potential has Fourier transform (e.g. Peebles 1980)

$$\Phi_2(\mathbf{k}, \eta) = -\frac{a(\eta)}{21 H_0^2 k^2} \int \frac{d^3 \mathbf{k}'}{(2\pi)^3} \Phi_1(\mathbf{k} - \mathbf{k}') \Phi_1(\mathbf{k}') (3 k^2 k'^2 + 7 k^2 \mathbf{k} \cdot \mathbf{k}' - 10 (\mathbf{k} \cdot \mathbf{k}')^2), \tag{2}$$

and $\Phi_1$ is the linear theory gravitational potential; $a(\eta) = (\eta/\eta_0)^2$ is the scale factor.

The three–point correlation function for points at three arbitrary angular separations $\alpha$, $\beta$ and $\gamma$ is given by the average product of temperature fluctuations in all possible three directions with those angular separations among them. The general expression is given by Gangui et al. (1994). In this paper, for simplicity, we will restrict ourselves to the *collapsed* case, corresponding to the choice $\alpha = \beta$ and $\gamma = 0$, that is one of the cases analysed for the *COBE* DMR data by Hinshaw et al. (1994, 1995) (the other is the *equilateral* one, $\alpha = \beta = \gamma$). The collapsed three–point correlation function of the CMB is given by

$$C_3(\alpha) \equiv \int \frac{d\Omega_{\hat{\gamma}_1}}{4\pi} \int \frac{d\Omega_{\hat{\gamma}_2}}{2\pi} \Delta T(\hat{\gamma}_1) \Delta T^2(\hat{\gamma}_2) \delta(\hat{\gamma}_1 \cdot \hat{\gamma}_2 - \cos\alpha). \tag{3}$$

For $\alpha = 0$, we recover the well–known expression for the skewness, $C_3(0)$. By expanding the temperature fluctuations in spherical harmonics $\Delta T(\theta, \varphi) = \sum_{\ell,m} a_{\ell m} Y_{\ell m}(\theta, \varphi)$, we can write the collapsed three–point function as

$$C_3(\alpha) = \frac{1}{4\pi} \sum_{\ell_1, \ell_2, \ell_3, m_1, m_2, m_3} P_{l_1}(\cos\alpha) a_{\ell_1 m_1} a_{\ell_2 m_2} a^*_{\ell_3 m_3} \mathcal{W}_{\ell_1} \mathcal{W}_{\ell_2} \mathcal{W}_{\ell_3} \mathcal{H}^{m_3 m_2 m_1}_{\ell_3 \ell_2 \ell_1}, \tag{4}$$

where $\mathcal{W}_\ell$ represents the window function of the particular experiment and we follow the notation in Gangui et al. (1994), defining

$$\mathcal{H}^{m_3 m_2 m_1}_{\ell_3 \ell_2 \ell_1} \equiv \int d\Omega_\gamma Y^{m_1}_{\ell_1}(\hat{\gamma}) Y^{m_2}_{\ell_2}(\hat{\gamma}) Y^{m_3 *}_{\ell_3}(\hat{\gamma}), \tag{5}$$

which have a simple expression in terms of Clebsh–Gordan coefficients. The mean angular bispectrum predicted by a given model can be obtained from

$$\langle a_{\ell_1 m_1} a_{\ell_2 m_2} a^*_{\ell_3 m_3} \rangle = \int d\Omega_{\hat{\gamma}_1} d\Omega_{\hat{\gamma}_2} d\Omega_{\hat{\gamma}_3} Y^{m_1*}_{\ell_1}(\hat{\gamma}_1) Y^{m_2*}_{\ell_2}(\hat{\gamma}_2) Y^{m_3}_{\ell_3}(\hat{\gamma}_3)$$
$$\langle \Delta T(\hat{\gamma}_1) \Delta T(\hat{\gamma}_2) \Delta T(\hat{\gamma}_3) \rangle \qquad (6)$$

[for a general discussion of the properties of the angular bispectrum see Luo (1994)]. The contribution from the Sachs–Wolfe term has been computed by Falk et al. (1993), Gangui et al. (1994) and Gangui (1994), accounting for primordial non–linearities due to inflaton self–interactions. The leading contribution coming from the Rees–Sciama effect is obtained from

$$\langle \Delta T(\hat{\gamma}_1) \Delta T(\hat{\gamma}_2) \Delta T(\hat{\gamma}_3) \rangle =$$
$$\frac{2}{9} T_0^3 \langle \Phi_1(\hat{\gamma}_1 \eta_0) \Phi_1(\hat{\gamma}_2 \eta_0) \int_{\eta_r}^{\eta_0} d\eta \frac{\partial}{\partial \eta} \Phi_2(\mathbf{x}, \eta) \Big|_{\mathbf{x} = \hat{\gamma}_3(\eta_0 - \eta)} \rangle$$
$$+ (\hat{\gamma}_1 \leftrightarrow \hat{\gamma}_3) + (\hat{\gamma}_2 \leftrightarrow \hat{\gamma}_3), \qquad (7)$$

with $T_0 = 2.726 \pm 0.01$ K the mean temperature of the CMB (Mather et al. 1994). At the same order in perturbation theory there are also non–vanishing contributions of the type $\frac{1}{27} \langle \Phi_1(\hat{\gamma}_1 \eta_0) \Phi_1(\hat{\gamma}_2 \eta_0) \Phi_2(\hat{\gamma}_3 \eta_0, \eta_r) \rangle$; compared to the non–local Rees–Sciama terms, these terms are however suppressed by about two orders of magnitude, because of both the $\eta$–dependence of $\Phi_2$ and the different numerical factors. Using the analytical expression for $\Phi_2$ from Eq. (2), a straightforward computation leads to

$$\langle \Delta T(\hat{\gamma}_1) \Delta T(\hat{\gamma}_2) \Delta T(\hat{\gamma}_3) \rangle =$$
$$-\frac{T_0^3}{189} \int \frac{d^3 \mathbf{k}_1}{(2\pi)^3} \int \frac{d^3 \mathbf{k}_2}{(2\pi)^3} \int_{\eta_r}^{\eta_0} d\eta \eta e^{i\mathbf{k}_1 \cdot \hat{\gamma}_1 \eta_0} e^{i\mathbf{k}_2 \cdot \hat{\gamma}_2 \eta_0} e^{-i(\mathbf{k}_1 + \mathbf{k}_2) \cdot \hat{\gamma}_3 (\eta_0 - \eta)}$$
$$\left( 3(k_1^2 + k_2^2) + 7(\mathbf{k}_1 + \mathbf{k}_2)^2 - \frac{10}{(\mathbf{k}_1 + \mathbf{k}_2)^2} \left( ((\mathbf{k}_1 + \mathbf{k}_2) \cdot \mathbf{k}_1)^2 + ((\mathbf{k}_1 + \mathbf{k}_2) \cdot \mathbf{k}_2)^2 \right) \right)$$
$$P_\Phi(k_1) P_\Phi(k_2) + (\hat{\gamma}_1 \leftrightarrow \hat{\gamma}_3) + (\hat{\gamma}_2 \leftrightarrow \hat{\gamma}_3), \qquad (8)$$

with $P_\Phi(k)$ the gravitational potential power spectrum. Using this expression and well–known integral relations for spherical harmonics, the three angular integrations in Eq. (6) can be performed and we obtain for the bispectrum

$$\langle a_{\ell_1 m_1} a_{\ell_2 m_2} a^*_{\ell_3 m_3} \rangle = -\frac{4 T_0^3}{63 \pi^2} \int d^3 \mathbf{k}_1 \int d^3 \mathbf{k}_2 \int_{\eta_r}^{\eta_0} d\eta \eta \sum_{j_1, j_2, n_1, n_2} i^{\ell_1 + \ell_2 - j_1 - j_2} \mathcal{H}^{n_1 n_2 m_3}_{j_1 j_2 \ell_3}$$

$$j_{\ell_1}(k_1\eta_0)j_{\ell_2}(k_2\eta_0)Y_{\ell_1}^{m_1*}(\hat{\mathbf{k}}_1)Y_{\ell_2}^{m_2*}(\hat{\mathbf{k}}_2)j_{j_1}(k_1(\eta_0-\eta))j_{j_2}(k_2(\eta_0-\eta))Y_{j_1}^{n_1}(\hat{\mathbf{k}}_1)$$
$$Y_{j_2}^{n_2*}(\hat{\mathbf{k}}_2)P_\Phi(k_1)P_\Phi(k_2)\left(\frac{20k_1^2k_2^2+8(\mathbf{k}_1\cdot\mathbf{k}_2)^2+14(k_1^2+k_2^2)\mathbf{k}_1\cdot\mathbf{k}_2}{k_1^2+k_2^2+2\mathbf{k}_1\cdot\mathbf{k}_2}\right), \quad (9)$$

where $j_\ell$ are spherical Bessel functions of order $\ell$. Now it is possible to make the integrations in $d\Omega_{\hat{\mathbf{k}}}$ and perform the summation over $m_i$ in Eq. (4), using relations among Clebsh–Gordan coefficients[1] to obtain for the collapsed three–point function

$$C_3(\alpha) = -\frac{A^2T_0^3}{63\pi^2 32\eta_0^6}\sum_{\ell_1\ell_2\ell_3}(2\ell_1+1)(2\ell_2+1)\mathcal{W}_{\ell_1}\mathcal{W}_{\ell_2}\mathcal{W}_{\ell_3}\langle\ell_1\ell_20\ 0|\ell_30\rangle^2 P_{\ell_1}(\cos\alpha)$$
$$\int_0^\infty dw_1 w_1^{n-3}\int_0^\infty dw_2 w_2^{n-3}\int_{\eta_r/\eta_0}^1 dz\frac{1-z}{z}J_{\ell_1+\frac{1}{2}}(w_1)J_{\ell_1+\frac{1}{2}}(w_1 z)$$
$$J_{\ell_2+\frac{1}{2}}(w_2)J_{\ell_2+\frac{1}{2}}(w_2 z)\left(10(w_1^2+w_2^2)+5\frac{(w_1^2-w_2^2)^2}{w_1 w_2}\ln\left|\frac{w_1-w_2}{w_1+w_2}\right|\right).(10)$$

One can readily see from the last expression the advantages of our approach: any primordial spectrum may be studied, as we have $P_\Phi(k) \equiv A(k\eta_0)^{n-4}$. We normalize its amplitude to the $Q_{rms-PS}$ value determined by *COBE* through $A/\eta_0^3 = (36/5)2^{1-n}4\pi Q_{rms-PS}^2 T_0^{-2}\Gamma^2(2-n/2)\Gamma(9/2-n/2)/(\Gamma(3-n)\Gamma(3/2+n/2))$. Moreover, the expansion in multipoles allows us to specify any chosen window function $\mathcal{W}_\ell$ as well as to subtract the desired multipole contributions from our expression, and thus match the different settings of various experiments (provided that we stay far enough from the Doppler peak). The integrals in Eq. (10) can be numerically evaluated to obtain the prediction for the collapsed three–point correlation function produced through the Rees–Sciama effect, to be measured by the particular experiment.

We turn now to the computation of the cosmic variance associated to the collapsed three–point correlation function. If the fluctuations in the CMB temperature were Gaussian, the mean value of the three–point correlation function over the ensemble of observers would be zero. However, as we are able to perform observations in just one particular sky, this prediction comes with a theoretical error bar, or "cosmic variance", that indicates the typical values expected for particular realizations of the Gaussian process. Only non–vanishing values larger than this can be interpreted as a signal of intrinsic non–random phases in the distribution of CMB temperature fluctuations.

---

[1] In particular, we use the identity $\sum_{m=-\ell}^{\ell}\langle\ell\ 2k\ m\ 0|\ell\ m\rangle = (2\ell+1)\delta_{k0}$.

The expected amplitude of the cosmic variance $\langle C_3^2(\alpha)\rangle$ at a particular angular scale can be computed from Eq. (3)

$$\langle C_3^2(\alpha)\rangle \equiv \int \frac{d\Omega_{\hat{\gamma}_1}}{4\pi} \int \frac{d\Omega_{\hat{\gamma}_2}}{2\pi} \int \frac{d\Omega_{\hat{\gamma}_3}}{4\pi} \int \frac{d\Omega_{\hat{\gamma}_4}}{2\pi} \delta(\hat{\gamma}_1 \cdot \hat{\gamma}_2 - \cos\alpha)\delta(\hat{\gamma}_3 \cdot \hat{\gamma}_4 - \cos\alpha)$$
$$\langle \Delta T(\hat{\gamma}_1)\Delta T^2(\hat{\gamma}_2)\Delta T(\hat{\gamma}_3)\Delta T^2(\hat{\gamma}_4)\rangle. \tag{11}$$

We assume that the multipole coefficients are Gaussian distributed random variables with angular spectrum $\langle a_{\ell_1}^{m_1} a_{\ell_2}^{m_2*}\rangle = \delta_{\ell_1\ell_2}\delta_{m_1m_2}(4\pi/5)Q_{rms-PS}^2 \mathcal{C}_{\ell_1}$, where

$$\mathcal{C}_\ell = \frac{\Gamma(\ell + n/2 - 1/2)\Gamma(9/2 - n/2)}{\Gamma(\ell + 5/2 - n/2)\Gamma(3/2 + n/2)}, \tag{12}$$

as it follows from $P_\Phi(k) \propto k^n$. Using standard combinatorial properties, we can compute $\langle C_3^2(\alpha)\rangle_{Gauss}$ from Eq. (11)

$$\langle C_3^2(\alpha)\rangle_{Gauss} = \frac{2Q_{rms-PS}^6}{125(4\pi)^{3/2}} \sum_{\ell_1\ell_2\ell_3} \mathcal{C}_{\ell_1}\mathcal{C}_{\ell_2}\mathcal{C}_{\ell_3} \mathcal{W}_{\ell_1}^2 \mathcal{W}_{\ell_2}^2 \mathcal{W}_{\ell_3}^2 (2\ell_1 + 1)(2\ell_2 + 1)$$
$$P_{\ell_1}(\cos\alpha)(P_{\ell_1}(\cos\alpha) + P_{\ell_2}(\cos\alpha) + P_{\ell_3}(\cos\alpha))\langle \ell_2\ell_3 0\, 0|\ell_1 0\rangle^2. \tag{13}$$

The last expression takes into account only the effect of the cosmic variance on the collapsed three–point function; one can easily modify it to allow for the experimental noise, by adding its contribution to the angular spectrum.

We can now apply our formalism to the *COBE* DMR measurements. We consider a Gaussian window function, $W_\ell \simeq \exp(-\frac{1}{2}\ell(\ell + 1)\sigma^2)$, with dispersion of the antenna–beam profile $\sigma = 3°.2$ (Wright et al. 1992). Figure 1 shows the expected collapsed three–point function arising from the Rees–Sciama effect, for three different values of the primordial spectral index $n = 0.7, 1, 1.3$. Note that all three types of perturbation spectra can be easily generated in the frame of inflation models (e.g. Mollerach, Matarrese & Lucchin 1994, and references therein). In all cases the perturbation amplitude has been normalized to two–year *COBE* DMR data, using the $\ell = 9$ multipole amplitude, $a_\ell = 9$ $\mu$K, according to the procedure proposed by Górski et al. (1994), which leads to $Q_{rms-PS} = 24.2, 19.5, 15.9$ $\mu$K, for $n = 0.7, 1, 1.3$, respectively. The top panel is obtained by subtracting from the map only the dipole contribution. In the central panel also the quadrupole and the octopole have been removed. In the bottom panel all the multipoles up to $\ell = 9$ have been subtracted. This procedure allows easy comparison with the recent analysis of the two–year data by the *COBE* team (Hinshaw et al. 1995). Figure 2 contains the corresponding plots for the *rms* collapsed three–point function obtained from our initially Gaussian fluctuation field, $\langle C_3^2(\alpha)\rangle_{Gauss}^{1/2}$. So far we have assumed full–sky coverage; the effect of partial coverage, due to the cut at Galactic latitude $|b| > 20°$, increases the cosmic variance and can be approximately taken into account by multiplying it by a factor 1.56 (Hinshaw et al. 1994). An analytic estimate (Srednicki 1993, Scott, Srednicki & White 1994) gives a somewhat smaller value, 1.23. Note that the subtraction of low order multipoles leads to a strong decrease both in the signal and in the cosmic variance. For all considered multipole subtractions and angular separations, the expected signal stays typically three orders of magnitude below the cosmic variance, which makes the effect undetectable. The results are found to be weakly dependent on the considered spectral index. For comparison, the amplitude of the collapsed three–point function produced by non–linearities in the inflaton dynamics (see Gangui et al. 1994) is about a factor of ten smaller than that arising from the Rees–Sciama effect. We can also evaluate the skewness parameter $\mathcal{S} \equiv \langle(\Delta T/T)^3\rangle/\langle(\Delta T/T)^2\rangle^2$, which has the advantage of being normalization (i.e. $Q_{rms-PS}$) independent for the considered effect; in the case, e.g., of only dipole subtraction and $n = 1$, $\mathcal{S} \approx -5$. Note that, in the same case, the heuristic estimate by Luo & Schramm (1993) leads to $\mathcal{S} \approx 18$, while the method by Munshi et al. (1995) yields a slight underestimate, $\mathcal{S} \approx -1$. The corresponding *rms* skewness parameter, arising from the cosmic variance of a scale–invariant Gaussian field, is $\mathcal{S}_{rms} \approx 1.3 \times 10^4 \, (Q_{rms-PS}/19.5 \, \mu K)^{-1}$.

One may try to increase the amplitude of the signal–to–cosmic variance ratio, by referring to experimental settings, such as FIRS (e.g. Ganga et al. 1993) and Tenerife (e.g. Hancock et al. 1994), which are more sensitive to intermediate angular scales. However, the little increase of this ratio is not enough to make it detectable; furthermore, one has to bear in mind that the Tenerife measurement is further affected by a large sample variance.

Finally, let us stress that our calculations apply to measurements of CMB anisotropies on scales large enough to be unaffected by the Doppler peak. The possible extension of this effect to smaller scales is discussed by Munshi et al. (1995), who however argue that also on those scales the cosmic variance largely overcomes the intrinsic non–Gaussian signal.

Italian MURST is acknowledged for financial support. A.G. acknowledges partial funding from the British Council/Fundación Antorchas under project 13145/1–0004.

# Figure captions

**Figure 1** The collapsed three–point function $C_3(\alpha)$, in $\mu K^3$ units, as predicted by the Rees–Sciama effect, vs. the angular scale $\alpha$. The dot–dashed line curves refer to the $n = 0.7$ spectral index, the solid line ones to $n = 1$ and the dashed line ones to $n = 1.3$. The top, middle and bottom panels represent the cases $l_{min} = 2$, 4, 10 respectively.

**Figure 2** The *rms* collapsed three–point function $\langle C_3^2(\alpha)\rangle_{Gauss}^{1/2}$, in $10^4$ $\mu K^3$ units, as arising from the cosmic variance of Gaussian fluctuations. Symbols are as in Fig. 1. Note that the top, middle and bottom panels, representing the cases $l_{min} = 2$, 4, 10 respectively, have different vertical scales.